# A Maturity Model for Public Administration Open Translation Data Providers

Núria Bel (Universitat Pompeu Fabra)

Mikel L. Forcada (Universitat d'Alacant)

Asunción Gómez-Pérez (Universidad Politécnica de Madrid)

**Abstract**

Any public administration that produces translation data can be a provider of useful reusable data to meet its own translation needs and the ones of other public organizations and private companies that work with texts of the same domain. These data can also be crucial to produce domain-tuned Machine Translation systems.

The organization's management of the translation process, the characteristics of the archives of the generated resources and of the infrastructure available to support them determine the efficiency and the effectiveness with which the materials produced can be converted into reusable data.

However, it is of utmost importance that the organizations themselves first become aware of the goods they are producing and, second, adapt their internal processes to become optimal providers. In this article, we propose a Maturity Model to help these organizations to achieve it by identifying the different stages of the management of translation data that determine the path to the aforementioned goal.

1. Introduction

Translated documents and their original source texts are becoming precious resources whose availability can bring about significant translation cost savings when large quantities of texts have to be translated. Existing translations are the raw material to feed *computer-assisted translation* (CAT) tools and *machine translation* (MT) systems to boost professional translation processes. An increasing demand for such resources has already been observed. However, the last survey by LT-Observatory[1] reports that while 83% of language service providers in Europe expects more language resources, in particular translation data, to become available in the next years, only a 15% is ready to pay for them.

Most likely, these expectations are based on two facts. On the one hand, the web contains large quantities of data that can be harvested and reused. The indexed web contains about 4,590 million pages,[2] with many being in multilingual sites, therefore constituting potential translation data. On the other hand, public administration organisms, such as the translation services of the European Commission, are producing translation resources that are made available for public reuse. In 2005, for instance, the EU institutions translated 2,861 million pages, according to a special report[3] written to assess the cost of translation in the EU. It represents about 70 % of the total EU translation volume.[4] Note that the expenditure for translation in 2005 amounted to 511 M€. The situation has surely changed because from 2005 to now, as Europe has moved from 15 to 24 languages.

---

[1] http://ltinnovate.blogspot.com.es/2016/04/the-future-of-language-resources-for.html
[2] According to www.worldwidewebsize.com at 15-05-16.

For the first source of translations, the web, a well founded-study conducted for the project QTLaunchPad[5] (Tsiavos et al. 2015) notices that although technically is not only possible, but even easy thanks to the availability of specialized crawlers for finding multilingual webs already available, reusing unauthorized crawled resources implies legal risks, mostly related with the copyrights of web published texts. The report recommends as mitigation strategy to ask for permission to every crawled web that has not published the license conditions about its contents. Arranz and Hamon (2012), in the framework of the PANACEA project, described, however, how costly it can be to contact every provider to ask a license: on average it took 176 days to negotiate the permission for reusing web contents with actual copyright holders. Other proposals for avoiding legal issues are presented in Forcada et al. (2016).

None of these problems arise for the resources provided by public administrations, if, as in the case the of the translation memories of the European Commission Directorate General for Translation (DGT), they are published as open data. Note that this resource is the most downloaded resource of the European Open Data Portal (which contains more than 8000 different datasets). Open data, by definition, guarantees availability, access, copyright and reuse permissions, in all conditions, also with commercial purposes. Moreover, Directive 2003/98/EC of the European Parliament and of the Council of 17 November 2003 on the reuse of information of the public sector ensures that this sharing policy extends to all public administrations of the EU Member States. In fact, the European Language Resource Coordination initiative is already working for Member States to realize this possibility and thus to contribute to the creation of *automated translation* services in the framework of the Connecting Europe Facility programs (CEF.AT). The proposal is to mine selected organizations to extract language resources from them.

The question that this invitation to mine language resources in public administration organizations raises is: what would be the cost of the exercise of identifying, gathering and making available these resources? Obviously the cost can be minimal if we talk about translation memories which are already compiled and formatted. However, the use of CAT tools in the administration might be not as spread as it might seem. For instance, the case in Spain is that according to the *Libro Blanco de la traducción y la interpretación institucional*[6] ('White Paper on Institutional Translation in Spain') only less than the 8% of the translators in these organizations works with CAT tools. What is, then, the cost of identifying, collecting and preparing translation data from the rest of potential translation data providers?

---

[3] Special Report No 9/2006 of the European Court of Auditors concerning translation expenditure incurred by the Commission, the Parliament and the Council ([2007/2077(INI)](#))
[4] Official Journal of the European Union, C 284/1 (http://eur-lex.europa.eu/legal-content/EN/TXT/?uri=CELEX%3A52006SA0009)
[5] http://www.qt21.eu/launchpad/system/files/deliverables/QTLP-Deliverable-4_5_1_0.pdf
[6] http://ec.europa.eu/spain/pdf/libro_blanco_traduccion_es.pdf

In the framework of the recently approved Spanish *Plan de Impulso de las Tecnologías del Lenguaje*[7] ('Language Technologies Support Program'), the *Recursos para las Tecnologías del Lenguaje* ('Resources for Language Technology') network of excellence[8] has been commissioned to conduct an study to identify Spanish public administration organizations that produce translation data to produce a catalogue of potential open data providers. An extensive survey is currently being carried out. In this paper we present the criteria to determine both what are the characteristics that actual translation related resources have and to what extent they can hamper its identification, collection and preparation, in order to identify the organizations that might become providers of a continuous supply of data. We suggest that these criteria can be structured as a path to follow for organizations to become regular translation data providers.

In what follows we address, in section 2, some technicalities about translation resources, in section 3, we propose a maturity model for organizations with a primary focus on Public Administration bodies. Some conclusions are provided in section 4.

## 2. Technicalities about language resources for CAT and SMT

Non-experts might have difficulties to realize the characteristics of the translation resources required to feed CAT tools or to build MT systems in particular. In what follow, we review most relevant aspects about these data that need to be taken into account when assessing the task of deciding about the benefit/cost balance of collecting resources.

The following translation data are of interest, but in what follows we mainly concentrate on documents either in the form of translation memories or corpora.

- Translation memories: linguistic databases that capture translations made by humans. They can be used to facilitate future translations tasks but also for training automated translation systems.
- Corpora: monolingual and multilingual corpora, comparable, aligned, parallel documents, etc.
- Lexica: monolingual and multilingual lists of words, multi-words, sentences, etc. in general or specific subject fields.
- Terminological resources: structured sets of concepts, with associated linguistic information in a specific subject field.

One of the musts for reusability is that data collections can be downloaded as a whole and freely processed, as stated by the Full Open Data definition,[9] including commercial purposes. There are "accessible" materials that are useless.[10]

Other required technical details to be taken into account are the following.

---

[7] http://www.agendadigital.gob.es/planes-actuaciones/Paginas/plan-impulso-tecnologias-lenguaje.aspx
[8] http://retele.linkeddata.es/ (founded by the Spanish Ministerio de Economía y Competitividad, TIN2015-68955-REDT)
[9] http://opendefinition.org/od/2.1/en/
[10] http://index.okfn.org/dataset/legislation/

## 2.1. Size

The magnitude of the resources required for building MT systems that translate with a quality that makes it worth their use for professional translation is a common topic in technical discussions. The easy answer is "the more, the best", which is initially the truth, but a threshold should be pointed out when assessing a potential repository.

Most standard MT datasets contain tens or hundreds of millions of words (Irvine & Callison-Burch, 2013). There are techniques to mitigate the need of large quantities of parallel text, but most often at the expense of resulting translation quality. As a reference of the magnitude we can take as a standard corpus the Common Crawl corpus (Smith et al. 2013) that in actual experiments like the ones by Song et al. (2014) contained 161 million words, 3,158,523 sentence pairs for the French–English language pair. Besides, again to guarantee quality, with a base of hundreds of millions, and very good data of a particular domain, about several hundreds of parallel sentences words (Pecina et al. 2014) can be enough to tune a system to achieve a good quality in a particular domain.

In the specialized literature, low-resource settings are considered to be those with parallel datasets of fewer than one million words. One million words are about 3,000 pages of parallel data: original and translated documents.

## 2.2. File format

Despite of the benefits that formats such as PDF might have for archiving and printing documents, documents in PDF cannot generally be directly processed. Note that in order to use them for CAT or MT system training, segmentation into sentences —or segments— is required for alignment. Plain, editable formats are thus required, and although conversion between formats is technically possible, there are still many problems when the source is a PDF file. Because the original source texts of PDF files were produced in other editable formats, it is highly recommended to use these sources for building a translation data repository.

## 2.3 Alignment.

Traditional document archiving protocols do not always require encoding different language versions. In the case they do, in order to process documents, the best practice is when the file name or path clearly identifies the relation between source and target documents and the language of the named document.

The effort for collecting documents and translations can significantly vary depending on the alignment information: a very disperse and undocumented or unlabelled collection of documents can demand a big effort, and therefore be completely prohibitive, in terms of effort to make it reusable.

The best option, however, is storing translation memories, as they constitute an already compiled collection of segments of the same topic related explicitly with their translations, and one that is aligned, in a particular, preferably, standard format.

## 2.4. Translation memories

A translation memory is a file where source texts and its translations are stored broken down into segments, which are aligned with their translations to form "translation units". Segments are mostly, but not only, sentences. Currently, CAT tools include the programs that produce translation memories when a human translator types the translation and when two documents are provided as source and target texts. These tools can also export in-built translation memories into a standard format, most commonly TMX: the *translation memory exchange* format.[11]

What are the contents in a translation memory file is decided by the user. A translation memory can store document-by-document translation units, or many documents belonging to a particular subject or domain can be stored into one single TMX file.

### 2.5. Metadata

Translation data packages need to be documented in order to be searchable by means of metadata. A standard metadata schema is already available[12] as suggested by the community and for these metadata to be assigned information has to be foreseen since the initial archiving step: languages, size, domain, character encodings as well as license of the resource or whether texts include contain private or confidential data are required as well as more clerical information: creation date, contact data of the responsible person, and whether there are associated documentation and resources.

## 3. Public Administration Open Translation Data Providers

Any public administration that produces translation data to meet its own translation needs is a potential provider of data for other organizations including private companies that work with texts of the same domain. The problem is that they are completely ignorant of the potentiality of the data they produce.

By producing translations (internally or by outsourcing), the administration produces language resources: data in the form of documents and their translations, translation memories, terminologies and bilingual or multilingual glossaries. The organization's management of the translation process, the characteristics of the archives of the generated resources and of the infrastructure available to support them determine the efficiency and the effectiveness with which the materials produced can be converted into reusable data. For this reason, here we focus on the organization itself, which can provide protocols for archiving documents (therefore, data) and requirements regarding the format, the associated information or metadata, the licenses and distribution restrictions because of confidentiality and privacy data. The existence of these protocols for the management of the translation process and the available infrastructure are the basis for developing objective criteria for assessing which organizations can be considered suppliers.

Here we propose a set of criteria for evaluating the potential cost of collecting and preparing these data. Criteria are provided as an organization maturity model to suggest guidelines to adapt the operations of administration offices for them to become translation data providers. This model offers, therefore, a clue about the elements that need to be improved by the organization for it to become a supplier of resources.

We take into account requirements based on the availability and readiness of the resources and their characteristics, as described in section 2. Note that, while any resource is potentially useful, the costs to effectively enable their reuse can be high, mostly because of the quantity of texts required, and, therefore, hinder the objective. Also it is worth to keep in mind that the investment in a common infrastructure (particular tools like anonymization tools, computer-assisted translation programs, assistance for licensing evaluation, treatment of private confidential data, etc.) may be a necessary requirement for an organization to effectively reach maturity stage 5, the highest in the scale that we explain below.

---

11  https://www.gala-global.org/tmx-14b
12  https://elrc-share.ilsp.gr/ELRC-SHARE_SCHEMA/v1.0/

For our *translation data provider maturity model* we have taken into account the *capability maturity model integration* by Sally Godfrey (2008), which is shown in Figure 1 and Aymerich and Carmelo (2009) report on translation services at the PanAmerican Health Organization.

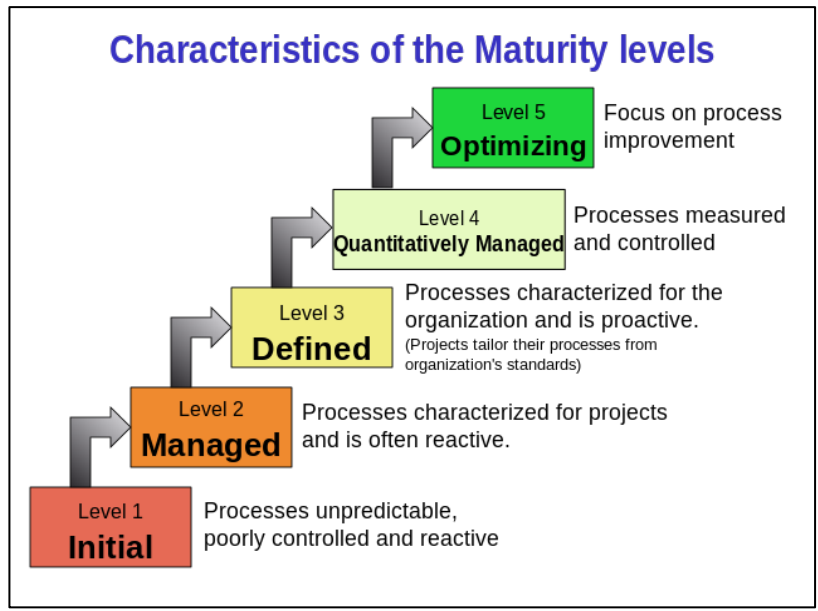

Figure 1: Capability Maturity Model Integration by Sally Godfrey (2008)

**Level 1: Initial**

At Level 1 or Initial ("Processes unpredictable, poorly controlled and reactive") one would find the organizations in which documents and their translations are not archived as a defined particular process. Translation data preservation is part of the general process of archiving documents, perhaps on paper or as a scanned document in PDF format, so that the relation between a document and its translation is not systematically registered and the registered copy is not processable. The organization has not been provided with tools that support the particular archiving of documents related to its translations.

At this level, linguistic resources could be reused, but only thanks to the competence and will of the people in the organization, but not as the result of to a proven process. The risk of considering these organizations as providers is that delivered materials might be inconsistent: it may be that not all translations are associated with the original document causing significant failures. In addition, organizations in maturity level 1 can commit themselves beyond their capabilities, or deliver resources of quality lower than expected, and therefore cannot guarantee the continuity of the process of compiling resources.

**Level 2: Managed**

At Level 2 ("Processes characterized for projects and often reactive"), the translation process has been given a particular space for archiving and it is managed by a protocol. Original texts and their translations are produced in editable formats: (.doc) Microsoft Office, OOXML (.docx), HTML, ODF (.odt), or plain text (.txt).

The *ad hoc* protocol only requires the identification (traceability) of documents and translations (e.g. there may have guidelines on how to name the folders and file documents with names where the only variation is an indication of the language of the document), but quality controls are not provided and there are no specific objectives as regards the quality of archived materials or file system.

The protocol manages the file that is in a space (common folder, Drive space, etc.) separately from other activity documentation file; however, the reuse of materials is not yet considered the ultimate aim of the process. It could be the case that translators use CAT, but each translator manages their own memories in their personal workstation, and even if they could share them, they would share them in an informal and unsystematic way.

**Level 3: Defined Objective**

At level 3 ("Processes characterized by the organization and is proactive: projects adapt their processes to organizational standards"), compiling and archiving of translation materials are integrated into the translation process, since the reuse of materials by using CAT programs is integrated into the translation process.

The document and translation archive in the form of a translation memory is planned from an early stage of the translation project and it is documented in the translation plan. Translation is planned relying on the availability of existing resources (reuse of translation memories of previous projects). The archiving protocol is reviewed from the experience that builds up in the organization.

A critical distinction between maturity level 2 and level 3 is the scope of the descriptions of processes, procedures and standards. At level 2 they may still vary for each instance or project. At level 3, the standards, processes and procedures are those used by the entire organization and expressed in a protocol that is tuned for each project. Another critical distinction is that at level 3 processes are described rigorously and are part of the content to be taken into account for staff training.

**Level 4: Quantitatively Managed**

After level 2 and 3, the organization has a technical and human infrastructure where the reuse of translations and related materials is considered an element of the translation process optimization.

In organizations in Level 4 ("Quantitative management: measured and controlled processes") compilation and archiving of translations for reuse in the form of translation memories are defined in detail as a process that is documented and measured, since it is perceived as a fundamental basis of the translation process; the resources generated are used to evaluate the costs of translation, to evaluate the productivity of the organization and to monitor improvements.

At Level 4 translation memories are stored and managed centrally, possibly in a version control system, which is updated manually when translators (or those responsible for managing subcontract translations) choose to do so. Memories, and their versions, are documented specifically with information concerning language, format, person responsible for the translation, segment size, creation date, domain and document type, character encoding and associated resources (terminologies, glossaries). Translation data are considered as an internal good, produced and controlled by the organization itself. However, in the protocol, the active search for possible external sources that can be incorporated to optimize the system is not contemplated.

**Level 5: Optimal**

An organization that has managed to reach this level has a support infrastructure for compiling and archiving translation memories and related materials. At level 5 ("Emphasis on process improvement"), the organization includes actions to improve its processes based on a quantitative perspective while technological means to achieve it are provided. The organization also foresees active tasks to search external resources (translation memories, terminologies, etc.) that may contribute to the efficiency of the translation process.

As at level 4, level 5 translation memories are stored and managed centrally with an automatic control system. The archive is automatically updated when translators (or those responsible for managing translation subcontracts) send translations.

Following a defined protocol, translation memories, and their versions, are documented with information on language, format, person responsible for the translation, size segments, creation date, domain and type of document, character encoding and resources (terminologies, glossaries) associated, but in level 5 this is done using specialized metadata.

Translation resources are considered as a good, produced and used by the organization itself but the organization is also aware of the possibility of third parties to reuse its memories in an example of reuse of public sector information. Therefore, the documentation of translation memories includes specific information about the possibility that they are published: content of private information (in some cases, to be able to publish the materials, they have been anonymized), confidential information. In addition, the organization has decided on a distribution method (open data, published in the corporate web with a restricted license or other possible solutions). This information is added to the documentation and metadata.

4. **Conclusions**

The reuse of translation resources has become popular with CAT tools that use translation memories: the compilation of translated material that is used to find —and to reuse— previously translated text segments. These translation memories are also the best material for training statistical machine translation systems.

Any public administration that produces translation data can be a potential provider of useful reusable data to meet its own translation needs and the ones of other public organizations and private companies that work with texts of the same domain.

The organization's management of the translation process, the characteristics of the archives of the generated resources and of the infrastructure available to support them determine the efficiency and the effectiveness with which the materials produced can be converted into reusable data.

However, it is of utmost importance that the organizations themselves first become aware of the goods they are producing and, second, adapt their internal processes to become optimal providers. In this article, we propose a maturity model to help these organizations to achieve optimal organizations by identifying the different stages of the management of translation data that determine the path to the aforementioned goal. The maturity model presented is still ongoing work; it is going to be validated in an extensive study to identify Spanish public administration organizations that produce translation data in order to produce a catalogue of potential open translation data providers.